\documentclass[aps,pra,reprint,groupedaddress, amsmath,amssymb]{revtex4-2}

\usepackage{graphicx}% Include figure files
\usepackage{todonotes}
\usepackage{xcolor}
\usepackage{hyperref}

\begin{document}

\newcommand\bra[1]{{\langle{#1}|}}
\newcommand\ket[1]{{|{#1}\rangle}}

%\preprint{}

\title{General and complete description of temporal photon correlations in cavity-enhanced spontaneous
parametric down-conversion}


\author{Chris M{\"u}ller}
\email[]{chris.mueller@physik.hu-berlin.de}
%\thanks{}
\author{Andreas Ahlrichs}
\author{Oliver Benson}
\affiliation{Nano-Optics and IRIS Adlershof, Humboldt-Universit{\"a}t zu Berlin, Berlin, Germany}

\date{\today}

\begin{abstract}
Heralded single photon sources are the most commonly used sources for optical quantum technology applications. There is strong demand for accurate prediction of their spectral features and temporal correlations with ever increasing precision. This is particularly important in connection with the intrinsically stochastic photon-pair generation process in heralded sources. Here we present a complete theoretical description of the temporal correlation of a signal-idler, signal-signal and signal-signal-idler coincidences of photons generated by continuous wave pumped cavity-enhanced spontaneous
parametric down-conversion. The theory excellently predicts the measurements, which has been experimentally confirmed in our setup utilizing single photon detectors with high temporal resolution. This enables us to resolve and analyze the multi-photon correlation functions in great detail.

\end{abstract}

%\keywords{}

\maketitle

Spontaneous parametric down-conversion (SPDC) is a widely used nonlinear process to generate photons in several quantum applications such as quantum cryptography \cite{Gisin2002}, quantum imaging \cite{Brida2010, Lemos2014} or quantum interfaces \cite{Polyakov2011, Bussieres2014}.  SPDC sources are also able to generate entangled photons \cite{Trojek2008,Stuart2013,Steinlechner2013} or photonic cluster states \cite{Istrati2019} to prove general physical concepts in loophole-free tests of Bell's theorem \cite{Shalm2015,Giustina2015,Vedovato2018} or to perform optical quantum information processing \cite{Pilnyak2017}. However, all of these applications are limited by the photon generation rate and the lack of the full a priori knowledge of the photon state. Unknown correlations or properties of the photon state quickly deteriorate any potential for quantum applications. 

The SPDC process generates two photons, called signal and idler, with strong spectral and temporal correlation. To enhance the generation rate for a specific wavelength, a nonlinear crystal can be placed inside a cavity, realizing a cavity-enhanced SPDC source \cite{Ou1999,Fekete2013,Monteiro2014,Scholz2009}. Moreover, the cavity geometry, the reflectivity of the mirrors and the free spectral range of the cavity determine the spectral structure and temporal correlation of the generated SPDC photons. The cavity provides a tool to tailor the parameters of the photon state, for example to match the emission of the SPDC photons to a specific atomic transition line \cite{Rambach2016,Scholz2009,Ahlrichs2016,Mottola2020}. Complementary to the spectral features, the control of temporal correlations is equally important. Examples are the use of multiple photon pair sources as resource for on-chip optical quantum information processing \cite{Wang2017,Wang2018} or the realization of time-bin entangled photon sources \cite{Kuklewicz2006,Rielaender2017}. Surprisingly, a theoretical description of the temporal correlation for signal-idler pairs has been described in literature only for special configurations \cite{Herzog2008} or has been given without any derivation, discussion or appropriate verification \cite{Kuklewicz,Xingxing,Lu2003}. Also, the contribution of unwanted uncorrelated photons, which occur in the intrinsically stochastic photon-pair generation in cavity-enhanced SPDC sources, has been only poorly addressed.

Here, we present a full theoretical description of the signal-idler correlation based on the spectral properties of a cavity-enhanced SPDC source \cite{Slattery2019}. Furthermore, we expand the theory to the signal-signal and signal-signal-idler correlation of two photon pairs. Since this includes multiple photon pairs in the cavity at the same time, where the first generated pair can stimulate another pair, we have to take stimulated and non-stimulated process into account. To verify the theoretical findings we use a type-II SPDC crystal placed in a triply-resonant (pump, signal and idler) cavity pumped far below threshold \cite{Ahlrichs2016}.

\section{Signal-Idler Correlation} \label{sec:signal_idler_correlation}
We first investigate the signal-idler correlation, which is crucial for applications of SPDC sources as heralded single photon sources. The second-order signal-idler correlation function $G^{(2)}_{si}(\tau)$ can be expressed with the electric field operators $\hat{E}^{(+/-)}$ \cite{Lu2003,Herzog2008}
\begin{align}
G^{(2)}_{si}(\tau)&=\bra{\Psi}E^{(-)}_{i}(t) E^{(-)}_{s}(t+\tau)E^{(+)}_{s}(t+\tau)E^{(+)}_{i}(t)\ket{\Psi} \label{eq:G2_expanded} \\
&=|E^{(+)}_{s}(t+\tau)E^{(+)}_{i}(t)\ket{\Psi}|^2, \label{eq:G2}
\end{align}
where $\ket{\Psi}$ is a two-photon state generated in an SPDC process
\begin{align}
\ket{\Psi}=\int{d\omega_s}\int{d\omega_i} \ \psi(\omega_s,\omega_i)  a^\dagger_s(\omega_s) a^\dagger_i(\omega_i) \ket{0,0}, \label{eq:Psi}
\end{align}
with the joint-spectral amplitude $\psi(\omega_s,\omega_i)$ \cite{Mosley2008,Zielnicki2018}. Here we discuss SPDC in the case where the bandwidth of the generated photons is much smaller than the central frequency. The electric field operator can then be approximated by \cite{Herzog2008}

\begin{align}
\hat{E}_{s/i}^{(+)}(t)\approx \int^\infty_{-\infty}{d\omega \,\hat{a}_{s/i}(\omega)\; e^{-i[\omega t-kz]}}, \label{eq:E}
\end{align}
where $\hat{a}_{si}(\omega)$ is the annihilation operator for the measured photon.
For a monochromatic pump, as in our case, the joint-spectral amplitude $\psi(\omega_s,\omega_i)$ can be simplified in terms of the phase-matching function $f(\omega_s,\omega_i)$ \cite{Branczyk:11}
\begin{align}
\psi(\omega_s,\omega_i) =\delta(\omega_p-\omega_s-\omega_i) f(\omega_s,\omega_i). \label{eq:f}
\end{align}
Plugging the definition of the E-field given in Eq.~\eqref{eq:E} and the wave function of a generated SPDC pair from Eq.~\eqref{eq:Psi} into the second-order signal-idler correlation function $G^{(2)}_{si}(\tau)$ defined in Eq.~\eqref{eq:G2}, leads to 

%\begin{widetext}
%\begin{multline}
%G^{(2)}_{si}(\tau_1)=\Biggl|\int^\infty_{-\infty}{d\omega_i \,\hat{a}_{i}(\omega_i) e^{-i[\omega_i t-k_i z_i]}} \int^\infty_{-\infty}{d\omega_s \,\hat{a}_{s}(\omega_s) e^{-i[\omega_s (t+\tau)-k_s z_s]}}  \\
%\int^\infty_{-\infty}{d\tilde{\omega}_s}\int^\infty_{-\infty}{d\tilde{\omega_i}} \ \delta(\omega_p-\tilde{\omega}_s-\tilde{\omega}_i) f(\tilde{\omega}_s,\tilde{\omega}_i) \hat{a}^\dagger_s(\tilde{\omega}_s)  \hat{a}^\dagger_i(\tilde{\omega}_i) \ket{0,0}\Biggl|^2. \label{eq:G2_unsimplified}
%\end{multline}
%\end{widetext}

\begin{multline}
G^{(2)}_{si}(\tau)=\Biggl|\int^\infty_{-\infty}{d\omega_i \,\hat{a}_{i}(\omega_i) e^{-i[\omega_i t-k_i z_i]}} \int^\infty_{-\infty}{d\tilde{\omega_i}} \\ 
\int^\infty_{-\infty}{d\omega_s \,\hat{a}_{s}(\omega_s) e^{-i[\omega_s (t+\tau)-k_s z_s]}}  \int^\infty_{-\infty}{d\tilde{\omega}_s} \\
\delta(\omega_p-\tilde{\omega}_s-\tilde{\omega}_i) f(\tilde{\omega}_s,\tilde{\omega}_i) \hat{a}^\dagger_s(\tilde{\omega}_s)  \hat{a}^\dagger_i(\tilde{\omega}_i) \ket{0,0}\Biggl|^2. \label{eq:G2_unsimplified}
\end{multline}
From the commutator $[\hat{a}_n(\omega),\hat{a}^\dagger_m(\tilde{\omega})]=\delta_{n,m} \delta(\omega-\tilde{\omega})$, we obtain a relation for the creation and annihilation operators \cite{Herzog2008}

\begin{align}
\hat{a}_n(\omega)\hat{a}_m^\dagger(\tilde{\omega})\ket{0}=\delta_{n,m}\delta(\omega-\tilde{\omega})\ket{0} \label{eq:commutator}
\end{align}
which can be used to simplify Eq. \eqref{eq:G2_unsimplified}, yielding

\begin{align}
G^{(2)}_{si}(\tau)=\biggl|\int^\infty_{-\infty}{d\omega_s}\  e^{-i\omega_s \tau} f(\omega_s,\omega_p-\omega_s)\biggl|^2. \label{eq:g2_signal_idler_simplified}
\end{align}
Eq.~\eqref{eq:g2_signal_idler_simplified} shows that the second-order signal-idler correlation function $G^{(2)}_{si}(\tau)$, for a monochromatic pump, is the absolute value squared of the Fourier transform of the phase-matching function $f(\omega_s,\omega_i)$. This equation was also given in \cite{Kuklewicz,Xingxing,Lu2003} but without derivation, verification or discussion. 

\begin{figure}
	\centering
		\includegraphics[width=0.400\textwidth]{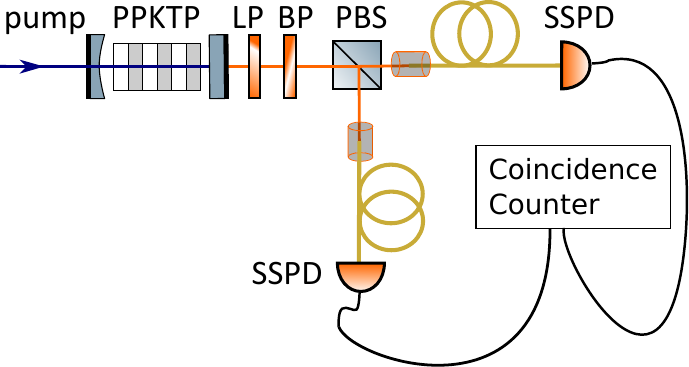}
	\caption{Setup for measuring the signal-idler correlation. The SPDC source consists of a 2\,cm PPKTP crystal, which is placed in a low finesse cavity to generate photons with 100\,MHz bandwidth \cite{Ahlrichs2016}. Residual pump light and non-phased-matched background are suppressed by a long-pass filter (LP) and a 1\,nm bandpass filter for 894\,nm (BP) after the cavity. Since the crystal is phase matched for type-II down-conversion, the signal and idler photons are split by a polarizing beam splitter (PBS) and collected into two separate single-mode fibers. SSPDs (Scontel) with a temporal resolution of 40\,ps for each detector are used to detect the signal and idler photons.}
	\label{fig:setup_signal_idler_correlation}
\end{figure}

To experimentally verify Eq.~\eqref{eq:g2_signal_idler_simplified}, we use a 2\,cm long periodically poled potassium titanyl phosphate (PPKTP) crystal with type-II phase-matching, which is placed in a triply resonant cavity and pumped far below threshold. The resonator has a finesse of about 8 for the pump photons and a higher finesse of about 15 for the signal/idler photons. The source was designed to emit degenerated signal and idler photons at the Cs D1 absorption line (894.3\,nm) with a bandwidth of 100\,MHz \cite{Ahlrichs2016}. A polarizing beam splitter (PBS) separates the generated signal and idler photons, which are coupled into two individual single-mode fibers (see Fig.~\ref{fig:setup_signal_idler_correlation}). Super conducting single-photon detectors (SSPDs) with low timing jitter are used to measure the photons. 

\begin{figure}
	\centering
		\includegraphics[width=0.5\textwidth]{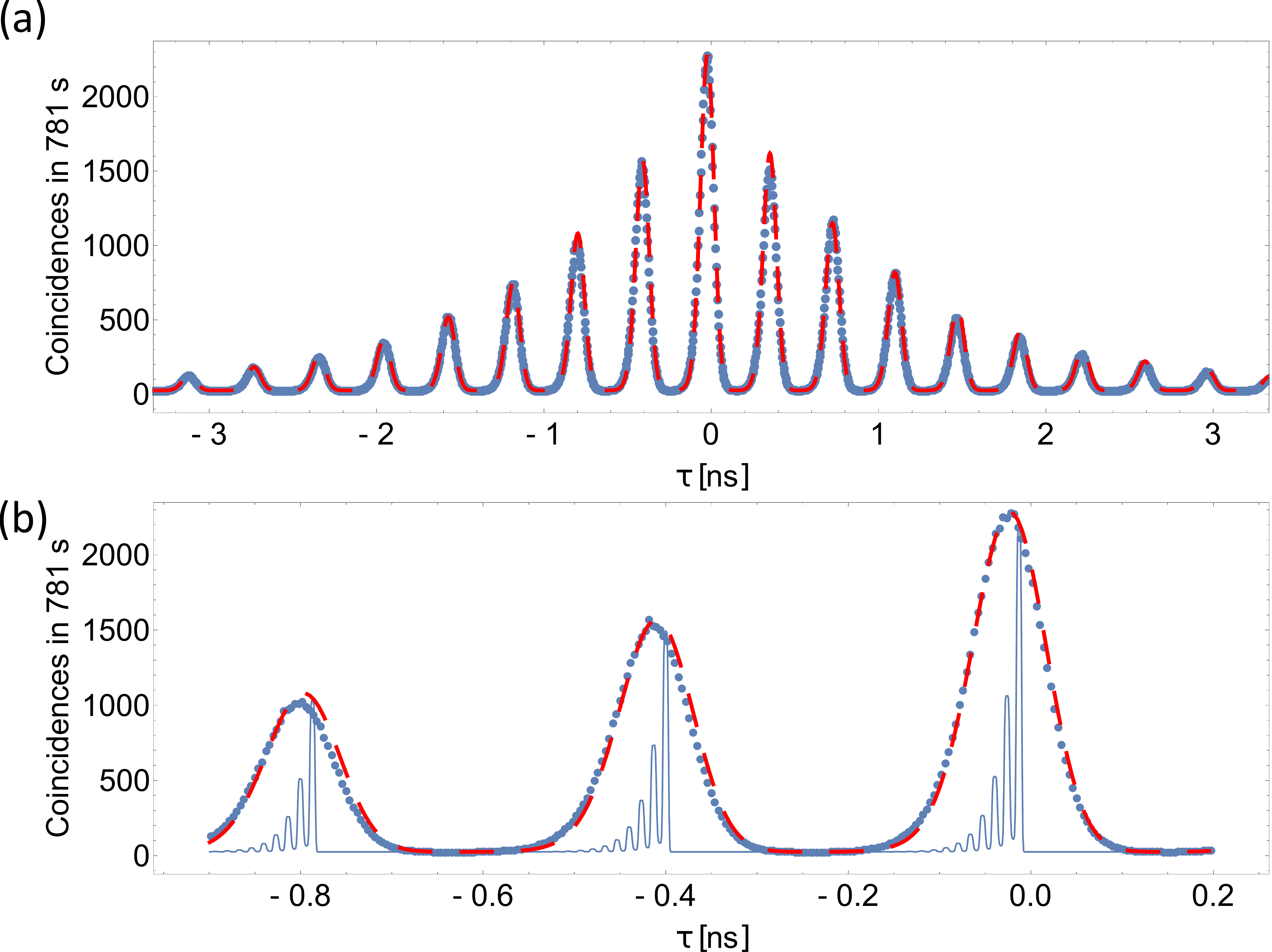}
	\caption{Measured and theoretical second-order signal-idler correlation. (a) Measured data (dots) and theoretical curve (dashed line) calculated with Eq.~\eqref{eq:g2_signal_idler_simplified}, using the phase-matching function of our source \cite{DissAhlrichs} convoluted with the overall temporal resolution of the setup. The distance between the peaks corresponds to the round-trip time of photons in the cavity. The exponential slope of the individual fringes is determined by the reflectivity of the outcoupling mirror. Negative correlation times correspond to the case where the signal photon left the cavity and was detected before the idler photon. (b) shows a zoom-in and a theoretical curve with a much higher temporal resolution (solid line). The substructures, which can not be resolved in the experiment, originate from the birefringence of the nonlinear crystal \cite{Herzog2008}.}
		\label{fig:signal_idler_correlation}
\end{figure}

The signal-idler coincidence measurement (dots in Fig.~\ref{fig:signal_idler_correlation}a) shows a comb structure where the height of individual peaks decrease with increasing detection time differences of signal and idler photon. This structure is caused by the cavity. The detection of one photon projects the other photon to a wave packet of several longitudinal cavity modes. This wave packet bounces back and forth between the cavity mirrors until eventually the photon escapes after an integer multiple of cavity round-trip times and is detected. The required phase-matching function to evaluate the theoretical finding of Eq.~\eqref{eq:g2_signal_idler_simplified} is determined by the cavity and crystal parameters. The number of measured coincidences depends mainly on the pump power, integration time and detection probability. However, the temporal correlation is independent of the number of measured events. Therefore, the theoretical curve is adjusted with respect to the maximum of the central peak at $\tau=0$. Furthermore, a constant offset was added to include uncorrelated background, e.g. generated by fluorescence inside the nonlinear crystal. The convolution of the theoretical curve with the instrument response function of our setup (dashed line in Fig.~\ref{fig:signal_idler_correlation}) fits exactly to our measurement. A zoom-in reveals that the theory predicts an interesting substructure in each peak (solid line in Fig.~\ref{fig:signal_idler_correlation}b) which, however, can not be resolved with our detectors. The different peaks of this finer structure occur, because the two photons of a pair accumulate an additional time difference each time they pass through the birefringent crystal \cite{Kuklewicz}. This constant time difference and the fixed separation of the signal and idler to their detectors, leads to an asymmetric correlation. When the idler photon is used as time reference, the substructure only occurs at negative correlation times.

%%%%%%%%%%%%%%%%%%%%%%%%%%%%%%%%%%%%%%%%%%%%%%%%%%%%%%%%%%%%%%%%%%%%%%%%%%%%%%%
\section{Signal-Signal Correlation}
In this part, we will address the signal-signal correlation. It characterizes the amount of pair bunching in the SPDC process. As we will see, there is a subtle influence of birefringence in this correlation, too. The second-order signal-signal correlation function is \cite{Foertsch2015}

\begin{align}
G^{(2)}_{ss}(\tau)&=\bra{\Psi}E^{(-)}_{s2}(t) E^{(-)}_{s1}(t+\tau)E^{(+)}_{s1}(t+\tau)E^{(+)}_{s2}(t)\ket{\Psi} \\ 
&=|E^{(+)}_{s1}(t+\tau)E^{(+)}_{s2}(t)\ket{\Psi}|^2. \label{eq:G2_signal_signal}
\end{align}
The wave function $\ket{\Psi}$ has to be expanded to two generated photon pairs, since we consider two signal photons from two different photon pairs. However, only signal photons in the same mode of the Fock state contribute relevantly to the signal-signal correlation. Otherwise, the second generated pair is not stimulated by the first generated pair and will be measured as a constant uncorrelated background. A delta function $\delta(\omega_s-\tilde{\omega}_{s})$ takes the stimulation of the second generated pair into account because it ensures that the two signal photons will be in the same mode. The idler photons will end up in the same mode automatically, because of energy and momentum conservation. The wave function $\ket{\Psi}$ is then
\begin{multline}
\ket{\Psi}=\int{d\omega_s}\int{d\omega_i}\int{d\tilde{\omega}_{s}}\int{d\tilde{\omega_{i}}} \quad \psi(\omega_s,\omega_i) \psi(\tilde{\omega}_{s},\tilde{\omega}_{i}) \\ 
\delta(\omega_s-\tilde{\omega_{s}}) \, a_s^\dagger(\omega_{s}) a_i^\dagger(\omega_{i}) a_s^\dagger(\tilde{\omega}_{s}) a_i^\dagger(\tilde{\omega}_{i}) \ket{0,0,0,0}. \label{eq:Psi4_1}
\end{multline}
Using the second-order signal-signal correlation function from Eq.~\eqref{eq:G2_signal_signal} with the definition of the E-field given by Eq.~\eqref{eq:E} and the wave function for two generated pairs defined in Eq.~\eqref{eq:Psi4_1} leads to
 
\begin{multline}
G^{(2)}_{ss}(\tau)= \Biggl| \int^\infty_{-\infty}{d\omega_{s1}} \int^\infty_{-\infty}{d\omega_{s2}}  \int^\infty_{-\infty}{d\omega'_{s}} \int^\infty_{-\infty}{d\omega'_{i}} \\  
\int^\infty_{-\infty}{d\tilde{\omega}_{s}} \int^\infty_{-\infty}{d\tilde{\omega_{i}}} \delta(\omega_p-\omega'_s-\omega'_i) f(\omega'_s,\omega'_i) \\ 
\delta(\omega_p-\tilde{\omega}_s-\tilde{\omega}_i) f(\tilde{\omega}_s,\tilde{\omega}_i) \delta(\omega'_s-\tilde{\omega}_{s}) \hat{a}_{s}(\omega_{s1}) \hat{a}_{s}(\omega_{s2}) \\
\hat{a}^\dagger_{i}(\omega'_i)  \hat{a}^\dagger_{s}(\omega'_s)  \hat{a}^\dagger_{s}(\tilde{\omega}_s) \hat{a}^\dagger_{i}(\tilde{\omega}_i)  
e^{-i(\omega_{s2} \tau)} \ket{0,0,0,0}\Biggl|^2.
\end{multline}
The ladder operators in this expression can be reduced with Eq.~\eqref{eq:commutator} to delta functions and the integrals can be simplified to 

\begin{multline}
G^{(2)}_{ss}(\tau)=\Biggl|  \int^\infty_{-\infty}{d\omega'_{s}} f(\omega'_{s},\omega_p-\omega'_{s})^2 \\
\hat{a}^{\dagger2}_{i}(\omega_p-\omega'_{s}) e^{-i(\omega'_{s} \tau)} \ket{0,0,0,0}\Biggl|^2.
\end{multline}
The remaining creation operator acts on the vacuum state and produces a factor of 2. Finally, the signal-signal correlation function is

\begin{multline}
G^{(2)}_{ss}(\tau)=\Biggl|  2 \int^\infty_{-\infty}{d\omega'_{s}} f(\omega'_{s},\omega_p-\omega'_{s})^2 \  e^{-i(\omega'_{s} \tau)}\Biggl|^2.
 \label{eq:g2_signal_signal_simplified}
\end{multline}
The expression is similar to the signal-idler correlation given by Eq.~\eqref{eq:g2_signal_idler_simplified}. However, the signal-signal correlation is the absolute value squared of the Fourier transform of the square of the phase-matching function.
 
\begin{figure}
	\centering
		\includegraphics[width=0.50\textwidth]{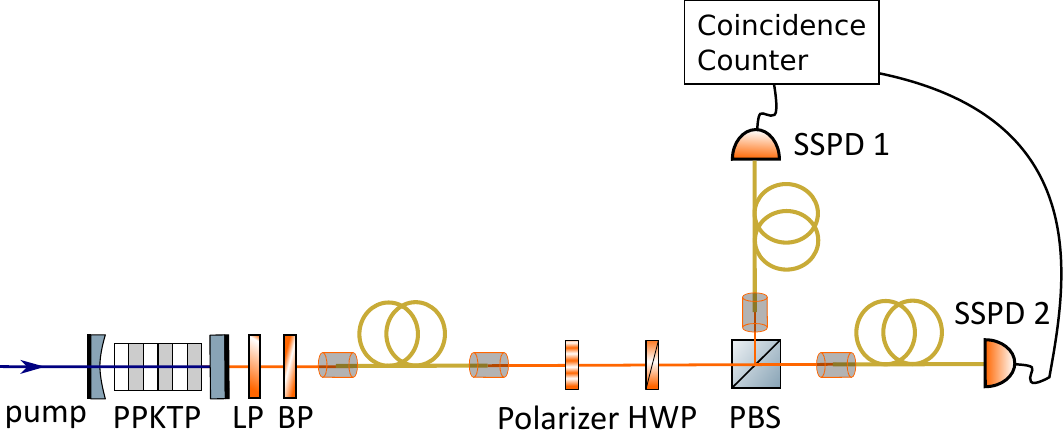}
	\caption{Setup for measuring the signal-signal correlation. The source is the same as for the signal-idler correlation, where signal and idler photons are orthogonally polarized. The emitted spectrum is tuned so that signal and idler photons have different frequencies. Therefore, the idler photons can be filtered out with a polarizer and a bandpass filter. The signal photons are send on a half-wave plate (HWP) and PBS combination to split them equally to two SSPDs (Scontel) with a temporal resolution of 40\,ps.}
	\label{fig:setup_signal_signal_correlation}
\end{figure}

\begin{figure}
	\centering
		\includegraphics[width=0.50\textwidth]{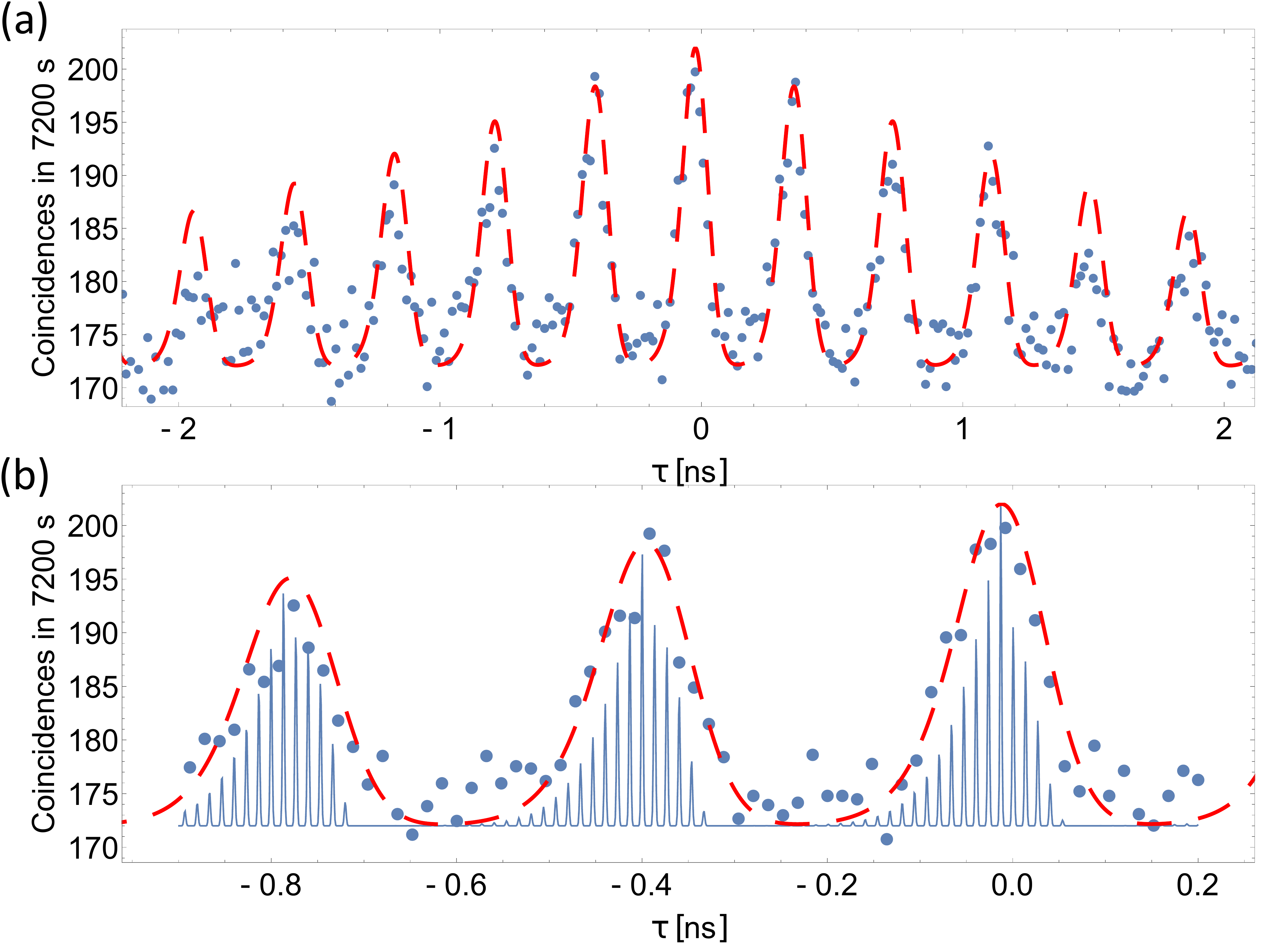}
	\caption{Measured (dots) and theoretical signal-signal correlation with a convolution of the overall temporal resolution of the setup (dashed line). Some of the idler photons passed the filter system and were detected. Hence, the theoretical line is a combination of 63\% signal-signal correlation with a residual contribution of 37\% signal-idler correlations as shown in Eq.~\eqref{eq:G2_ss_si}. (a) The correlation has peaks as for the signal-idler correlation, where the distance corresponding to the round-trip time. (b) A zoom-in and a much higher temporal resolution assumed in the theoretical curve (solid line) reveals a substructure again, which arises by the combination of the idler and signal probability to stimulate the second pair. The substructure has positive and negative contributions since both detectors have an equal probability to detect the first photon. }
		\label{fig:signal_signal_correlation}
\end{figure}

To verify Eq.~\eqref{eq:g2_signal_signal_simplified} experimentally, we use the same cavity-enhanced SPDC source as for the signal-idler correlation (see Sec.~\ref{sec:signal_idler_correlation}). In the experiment, we want to measure the temporal correlation of the signal photons. To achieve that, the nonlinear crystal was temperature tuned so that idler and signal spectra are no longer degenerate. This enables spectral filtering by a 1\,nm bandpass filter in addition to polarization filtering to block the idler photons. This is necessary since the suppression of just a polarization filter is insufficient to record the signal-signal correlation function. The signal photons pass the filtering system and are directed to a half-wave plate (HWP) which rotates their polarization to diagonal so that the PBS splits them equally (see Fig.~\ref{fig:setup_signal_signal_correlation}). Again, the signal-signal correlation shows a comb structure (dots in Fig.~\ref{fig:signal_signal_correlation}a) with a decreasing probability for longer time differences $\tau$ between detected photons. The position of the peaks in this measurement is, analogous to the signal-idler case, caused by the round-trip time in the cavity. In spite of the spectral and polarization filtering, there is still a contribution of residual signal-idler correlations in the measurement. In order to account for these contributions, we use the signal-signal correlation given in Eq.~\eqref{eq:g2_signal_signal_simplified} in combination with residual signal-idler correlation of Eq.~\eqref{eq:g2_signal_idler_simplified}
\begin{align}
G^{(2)}_{ss/si}(\tau)=a \cdot G^{(2)}_{ss}(\tau)+ b \cdot G^{(2)}_{si}(\tau) \label{eq:G2_ss_si}
\end{align}
with the parameters a=0.63 and b=0.37 (dashed line in Fig.~\ref{fig:signal_signal_correlation}a). The background of this measurement is higher than for the signal-idler case, due to uncorrelated signal photons from non-stimulated pairs.

A closer look at the theoretical curve without a temporal convolution shows a substructure in the peaks (line in Fig.~\ref{fig:signal_signal_correlation}b). Both detectors have an equal probability to detect the first photon, leading to a positive and negative part of the substructure. The signal and idler photon of the first pair have a defined temporal correlation and are able to stimulate the second pair. If just the signal photon from the first pair would be able to stimulate the second pair, we would expect a single peak without substructure. However, the idler photon from the first pair is able to stimulate a pair as well so that we obtain a superposition of both probability amplitudes, leading to a subtle birefringence-induced substructure in the signal-signal correlation.

%%%%%%%%%%%%%%%%%%%%%%%%%%%%%%%%%%%%%%%%%%%%%%%%%%%%%%%%%%%%%%%%%%%%%%%%%%%%%%%
\section{Signal-Signal-Idler Correlation}
\begin{figure}
	\centering
		\includegraphics[width=0.50\textwidth]{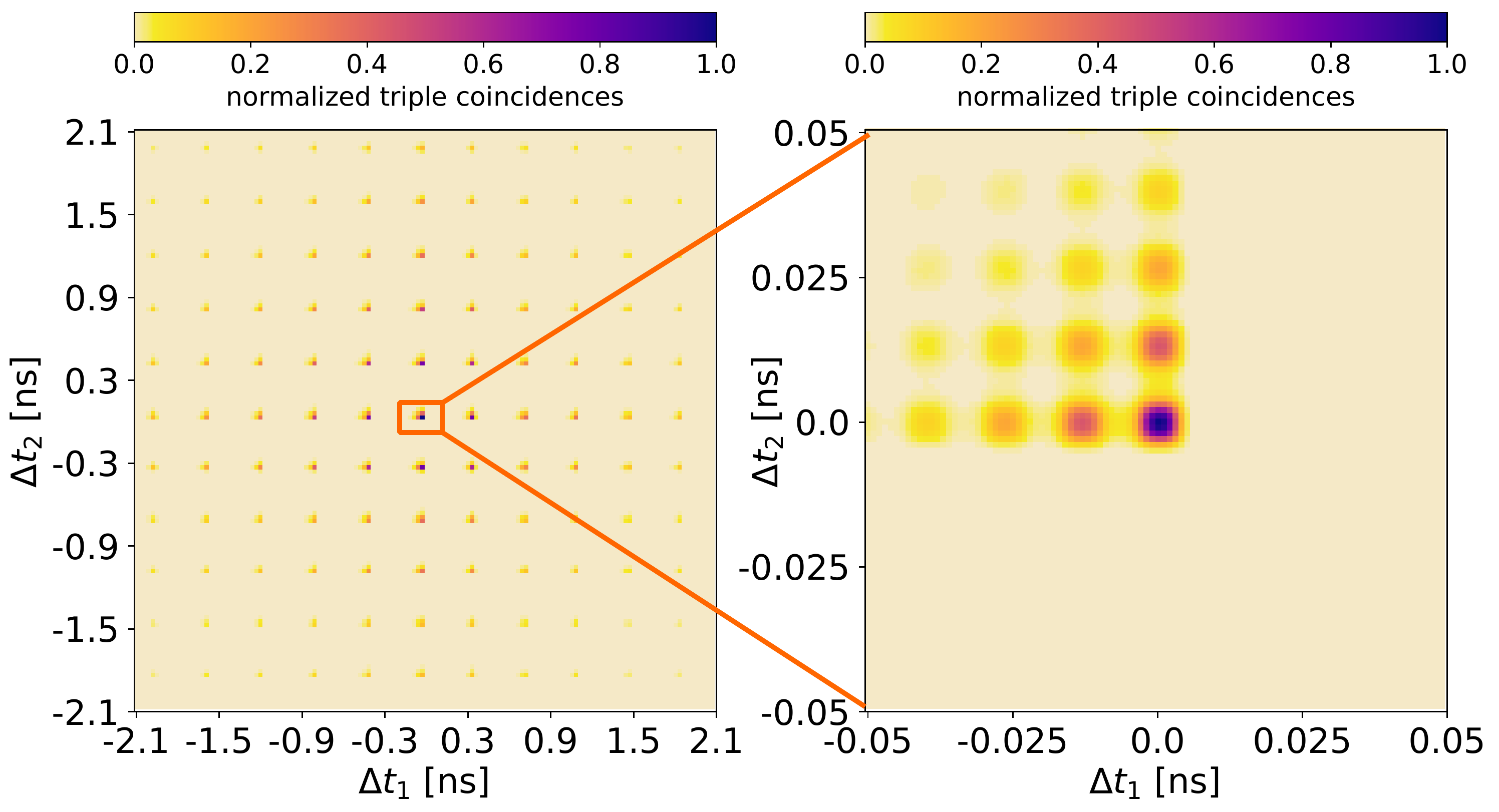}
	\caption{Second-order signal-signal-idler correlation of two simultaneously generated photon pairs. Calculated with Eq.~\eqref{eqn:two_dimensions} for a temporal resolution of 1\,ps (left picture is binned to 30 ps afterwards for improved illustration). $\Delta$t$_1$ ($\Delta$t$_2$) is the time difference between a detected signal photon~1 (signal photon~2) and the detection of the idler photon. The highest probability of a triple coincidence is when all involved photons leave the cavity together. This probability decreases with each additional round-trip of one of the photons. Zooming in shows a substructure of the correlation. This is caused by birefringence in the crystal for the signal and idler photons, which are polarized orthogonal to each other. The substructure is similar to the substructure in the signal-idler correlation (see Fig.~\ref{fig:signal_idler_correlation}).}
		\label{fig:m3000p3000_resolution_1ps_for_G00}
\end{figure}

Finally, we discuss the signal-signal-idler correlation. This correlation addresses the question how multi-photon detection events are influenced by uncorrelated photons also generated in the stochastic pair generation in the SPDC process. The three-photon signal-signal-idler correlation is the simplest, yet instructive example for multi-photon correlations. The second-order signal-signal-idler correlation can be described analogously by \cite{signal-signal-idler}
\begin{align}
G^{(2)}_{ssi}(\tau_1,\tau_2)=|E^{(+)}_{s1}(t+\tau_1)E^{(+)}_{s2}(t+\tau_2)E^{(+)}_{i}(t)\ket{\Psi}|^2. \label{G2_3_photons}
\end{align}
Since we consider three involved photons, the wave function $\ket{\Psi}$ has to take two generated photon pairs into account

\begin{multline}
\ket{\Psi}=\int{d\omega_s}\int{d\omega_i}\int{d\tilde{\omega}_{s}}\int{d\tilde{\omega_{i}}} \quad
 \psi(\omega_s,\omega_i) \psi(\omega_{\tilde{s}},\omega_{\tilde{i}}) \\ 
a_s^\dagger(\omega_{s}) a_i^\dagger(\omega_{i}) a_s^\dagger(\tilde{\omega}_{s}) a_i^\dagger(\tilde{\omega}_{i}) \ket{0,0,0,0}. \label{eq:Psi4}
\end{multline}
Using the second-order signal-signal-idler correlation of Eq.~\eqref{G2_3_photons} with the definition of the E-field of Eq.~\eqref{eq:E} and the wave function from Eq.~\eqref{eq:Psi4} leads to 

\begin{multline}
G^{(2)}_{ssi}(\tau_1,\tau_2)=
\Biggl| \int^\infty_{-\infty}{d\omega_{s1}} \int^\infty_{-\infty}{d\omega_{s2}} \int^\infty_{-\infty}{d\omega_{i}} \int^\infty_{-\infty}{d\omega'_{s}} \\ 
\int^\infty_{-\infty}{d\omega'_{i}} \int^\infty_{-\infty}{d\tilde{\omega}_{s}} \int^\infty_{-\infty}{d\tilde{\omega_{i}}}    \delta(\omega_p-\omega'_s-\omega'_i) f(\omega'_s,\omega'_i) \\
\delta(\omega_p-\tilde{\omega}_s-\tilde{\omega}_i) f(\tilde{\omega}_s,\tilde{\omega}_i) \hat{a}_{s}(\omega_{s1}) \hat{a}_{s}(\omega_{s2}) \hat{a}_{i}(\omega_i) \\ \hat{a}^\dagger_{i}(\omega'_i)  \hat{a}^\dagger_{s}(\omega'_s)  \hat{a}^\dagger_{s}(\tilde{\omega}_s) \hat{a}^\dagger_{i}(\tilde{\omega}_i)  
e^{-i(\omega_{s1}\tau_1+\omega_{s2} \tau_2)} \ket{0,0,0,0}\Biggl|^2.
\end{multline}
This expression can be simplified further with Eq.~\eqref{eq:commutator} and solving the integrals with the delta functions to

\begin{multline}
G^{(2)}_{ssi}(\tau_1,\tau_2)= \\ \Biggl| \int^\infty_{-\infty}{d\omega_{s1}} \int^\infty_{-\infty}{d\omega_{s2}} \ \hat{a}^\dagger_{i}(\omega_p-\omega_{s2})  f(\omega_{s2},\omega_p-\omega_{s2}) \\
 f(\omega_{s1},\omega_p-\omega_{s1}) \
  e^{-i(\omega_{s1}\tau_1+\omega_{s2} \tau_2)} \ket{0,0,0,0}\Biggl|^2.
\end{multline}
The action of the remaining annihilation operator $\hat{a}^\dagger_{i}(\omega_p-\omega_{s2})$ on the vacuum state, can be evaluated and results in

\begin{multline}
G^{(2)}_{ssi}(\tau_1,\tau_2)=\biggl|\int^\infty_{-\infty}{ d\omega_{s1} \ f(\omega_{s1},\omega_p-\omega_{s1}) e^{-i\omega_{s1}\tau_1}} \\ \int^\infty_{-\infty}{ d\omega_{s2} \ f(\omega_{s2},\omega_p-\omega_{s2}) e^{-i\omega_s\tau_2}}\biggl|^2. \label{eqn:two_dimensions}
\end{multline}
Eq.~\eqref{eqn:two_dimensions} is analog to Eq.~\eqref{eq:g2_signal_idler_simplified} but has two independent integrals, one for each photon pair. Eq.~\eqref{eqn:two_dimensions} describes the temporal behavior of two simultaneously generated photon pairs, which can be in different modes (see Fig.~\ref{fig:m3000p3000_resolution_1ps_for_G00}). There are only certain time points where a triple coincidence can occur. The time intervals depend on the round-trip time of the cavity. The highest triple coincidence probability appears when all three detected photons leave the cavity at the same time. Each additional reflection of a photon at the outcoupling mirror reduces the coincidence probability, as in the signal-idler correlation. A zoom-in reveals a substructure as in the signal-idler case (see Fig.~\ref{fig:signal_idler_correlation}) which arises from the slightly different round-trip times of signal and idler due to the birefringence. The substructure has only negative contributions, since the idler photon has a longer round-trip time in the cavity and its detection defines $\tau=0$.

\begin{figure}
	\centering
		\includegraphics[width=0.40\textwidth]{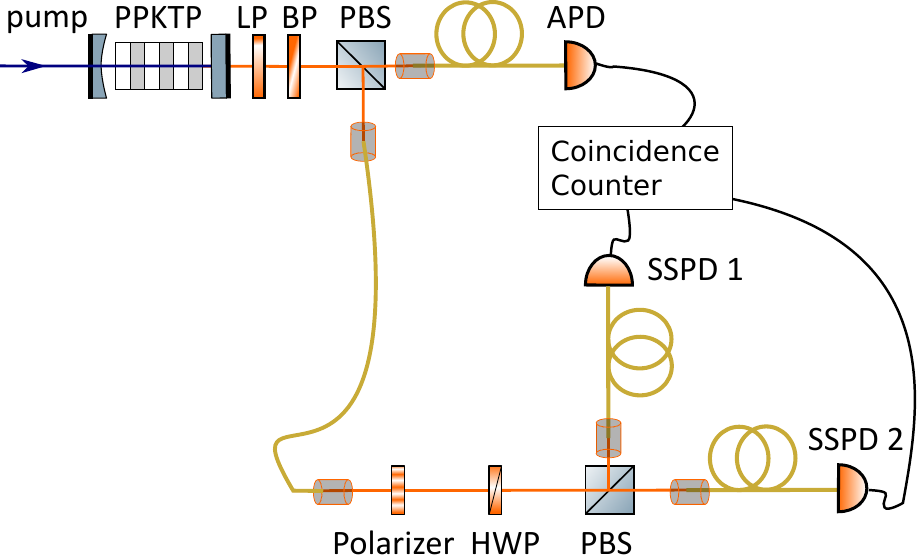}
	\caption{Setup for measuring the second-order signal-signal-idler correlation. The source is identical to the one used to measure the signal-idler correlation (see Fig.~\ref{fig:setup_signal_idler_correlation}). The signal and idler photons are split on a PBS. The idler photons are detected by a fiber-coupled avalanche photo diode (APD) with a timing resolution of 60\,ps. The signal photons are collected with a polarization maintaining fiber and send to a polarizer to suppress remaining idler photons. The transmitted signal photons pass a HWP and a PBS. The HWP rotates the polarization of the signal photons to diagonal so that the following PBS acts like a 50/50 beam splitter. The signal photons are detected individually with the two SSPDs with a temporal resolution beter than 40\,ps (Scontel).   }
		\label{fig:setup_signal_signal_idler}
\end{figure}

To verify the theoretical prediction in the measurement, we have to expand the setup for measuring the signal-idler correlation by an additional polarization maintaining single-mode fibers, a HWP and a PBS to split up the signal photons with a 50:50 chance on two SSPDs, as shown in Fig.~\ref{fig:setup_signal_signal_idler}. We use an avalanche photodiode (APD) with high temporal resolution (Micro Photon Devices, 60\,ps resolution) to detect the idler photons.  

In order to verify the theoretical prediction, Eq.~\eqref{eqn:two_dimensions} has to be adapted to the experimental setup. Since the second signal photon is uncorrelated with the heralded one, it can be generated later or earlier as well. Therefore, we have to take all possible generation times into account. If the uncorrelated photon is detected at SSPD~1, we have to modify Eq.~\eqref{eqn:two_dimensions} to

\begin{align}
\widetilde{G}^{(2)}_{ssi}(t_1,\tau_1)=\int_{t_{1min}}^{t_{1max}} dt_1 \left( G^{(2)}_{ssi}(\tau_1+ t_1,\tau_2)\right), \label{eq:background_t1}
\end{align}
where $t_1$ is the generation time difference of the two pairs, $t_{1min}$ and $t_{1max}$ are defining the minimum and maximum correlation times that we consider.

\begin{figure}
	\centering
		\includegraphics[width=0.500\textwidth]{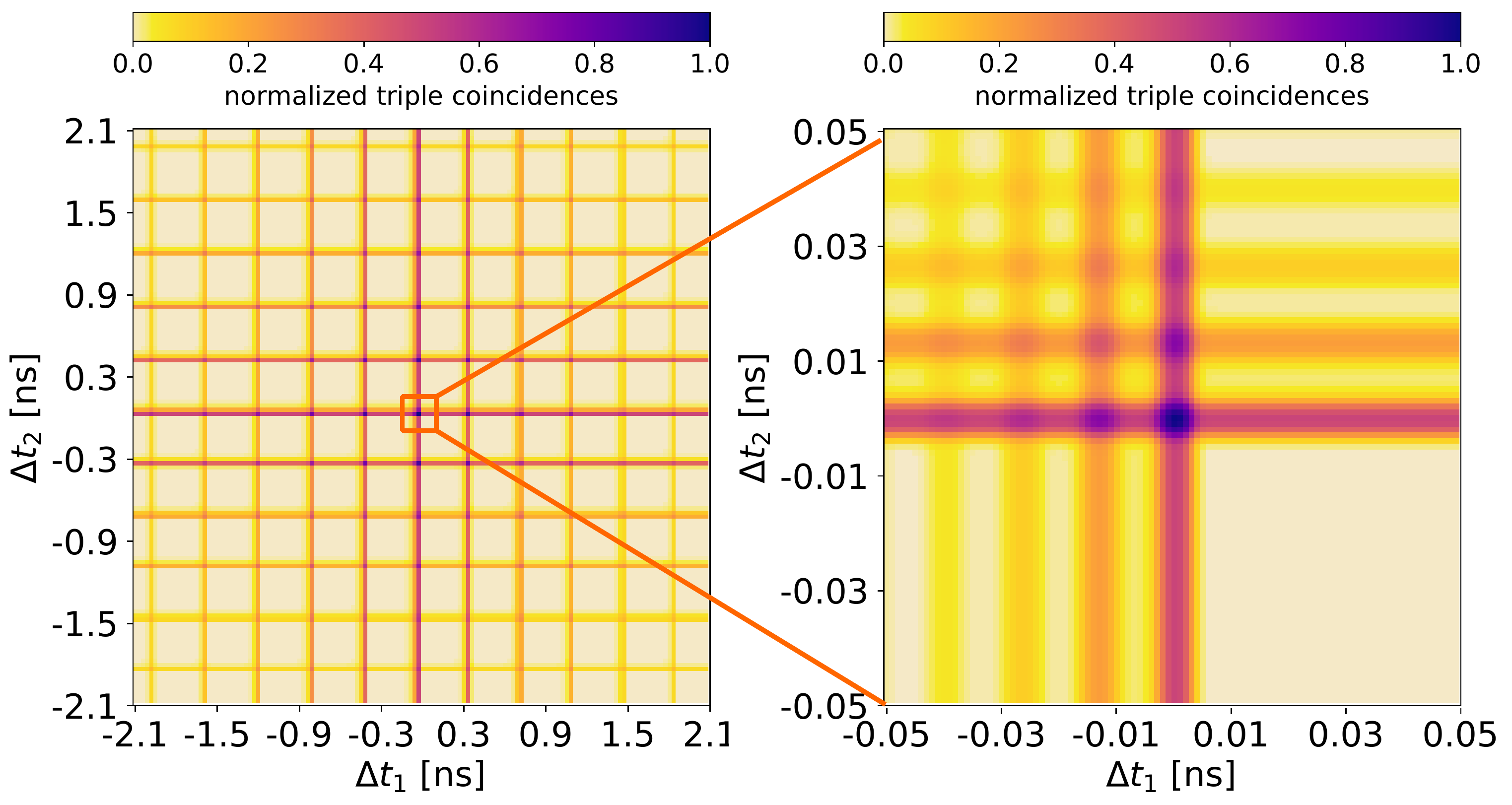}
	\caption{Theoretical second-order signal-signal-idler correlation taking uncorrelated photon pairs into account, calculated with a resolution of 1\,ps (left picture is binned to 30\,ps for improved illustration). The time difference $\Delta$t$_1$ ($\Delta$t$_2$) describes time different between a detected photon at the heralding APD and SSPD~1 (SSPD~2). The peak positions of the coincidences are equal to the case of simultaneously generated photons. Additionally, stripes can be seen which are caused by uncorrelated signal photons. At least one of the two signal photons has a time correlation to the detected idler photon, allowing the detection of a triple coincidence only at certain times. The uncorrelated signal photons are generated at a different time then the detected idler photon, resulting in the constant stripes. }
		\label{fig:dnum600p600_1200_summation_resolution_10ps_avby10}
\end{figure}

Uncorrelated photons can not only be detected at SSPD~1 but also at SSPD~2. Taking both possibilities into account, leads to

\begin{multline}
\widetilde{G}^{(2)}_{ssi}(t_1,t_2,\tau_1,\tau_2)=\int_{t_{1min}}^{t_{1max}} dt_1 \left( G^{(2)}_{ssi}(\tau_1+ t_1,\tau_2)\right) \\ 
+ \int_{t_{2 min}}^{t_{2 max}} dt_2\left( G^{(2)}_{ssi}(\tau_1,\tau_2+ t_2)\right) -G^{(2)}_{ssi}(\tau_1,\tau_2) , \label{eqn:two_dimensions_sum}
\end{multline}
where we subtract $G^{(2)}_{ssi}(\tau_1,\tau_2)$ because the simultaneous generated pairs would be counted twice otherwise. Fig.~\ref{fig:dnum600p600_1200_summation_resolution_10ps_avby10} shows a evaluation of Eq.~\eqref{eqn:two_dimensions_sum}, revealing additional stripes in comparison to the plot of two simultaneously generated photon pairs. These stripes occur when the two detected photon pairs have no time correlation to each other. However, one signal photon has always a temporal correlation to the detected idler photon, leading to the vanishing triple coincidence probability at certain times. It can happen that the idler photon of the first generated pair is not detected due to losses, but the idler of a later generated pair, leading to negative correlation times.

\begin{figure}
		\includegraphics[width=.390\textwidth]{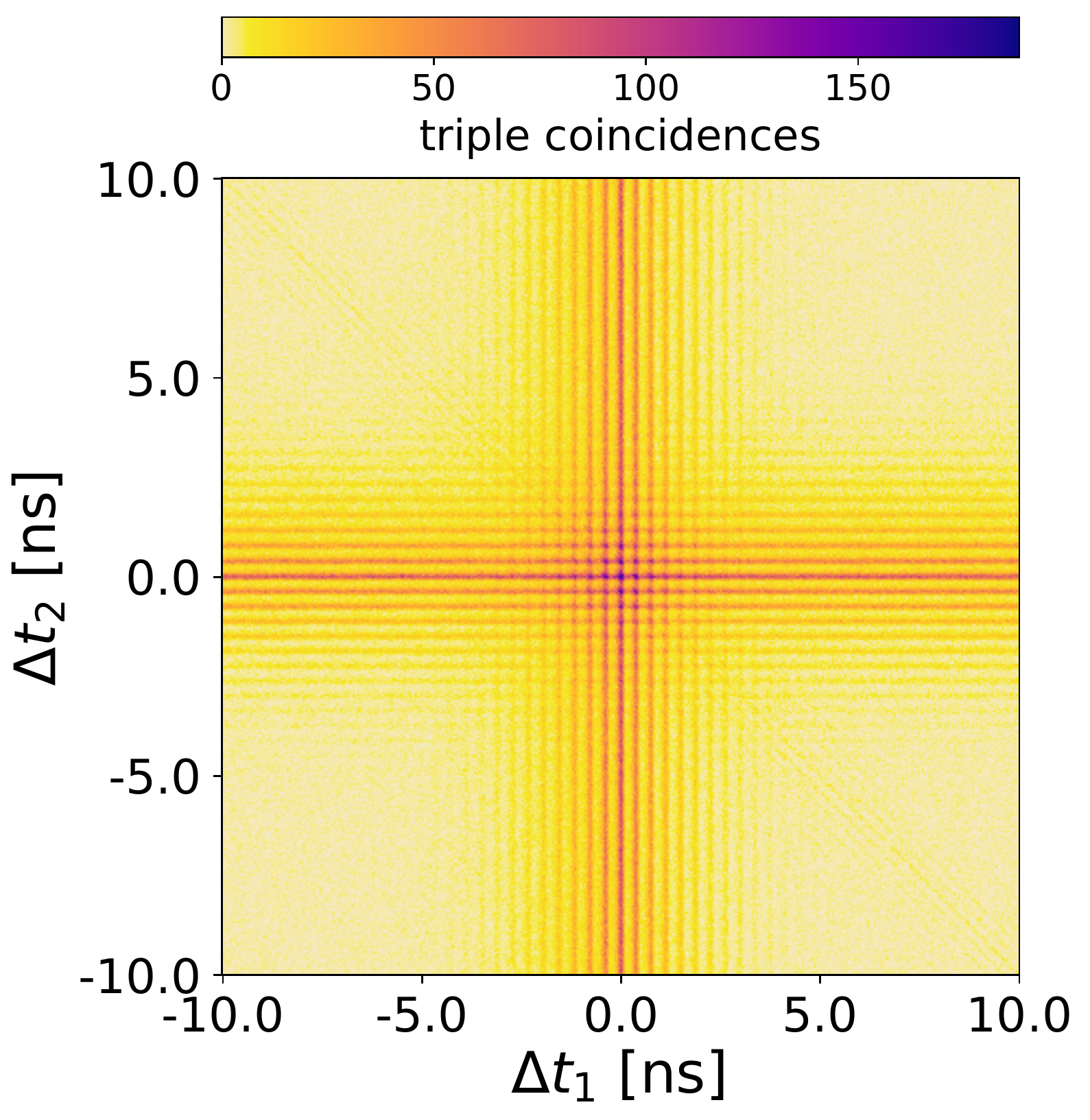}
	\caption{Measured second-order signal-signal-idler correlation with a binning resolution of 50\,ps. The correlation resembles a cross with the highest coincidence rates in the middle. The triple coincidence probability decreases rapidly with increasing time difference as expected from the theory (see Fig.~\ref{fig:dnum600p600_1200_summation_resolution_10ps_avby10}). The stripes are caused by signal photons, which where not generated at the same time as the detected idler photons. The stripes are constant over the entire measurement as predicted by the theory.}
		\label{fig:50ps_binning_measurement}
\end{figure}

Fig.~\ref{fig:50ps_binning_measurement} shows the measured second-order signal-signal-idler correlation for a correlation time window of 20\,ns.  It can clearly be seen, that the highest probability of a triple coincidence is detected when all photons are leaving the cavity simultaneously. Additional round-trips decreases the triple coincidence probability significantly. The stripes of the uncorrelated photons are constant over the entire measurement. A comparison with the theoretical prediction (see Fig.~\ref{fig:dnum600p600_1200_summation_resolution_10ps_avby10}) shows that the timing resolution was sufficient to resolve most features. 

\begin{figure}
		\includegraphics[width=0.500\textwidth]{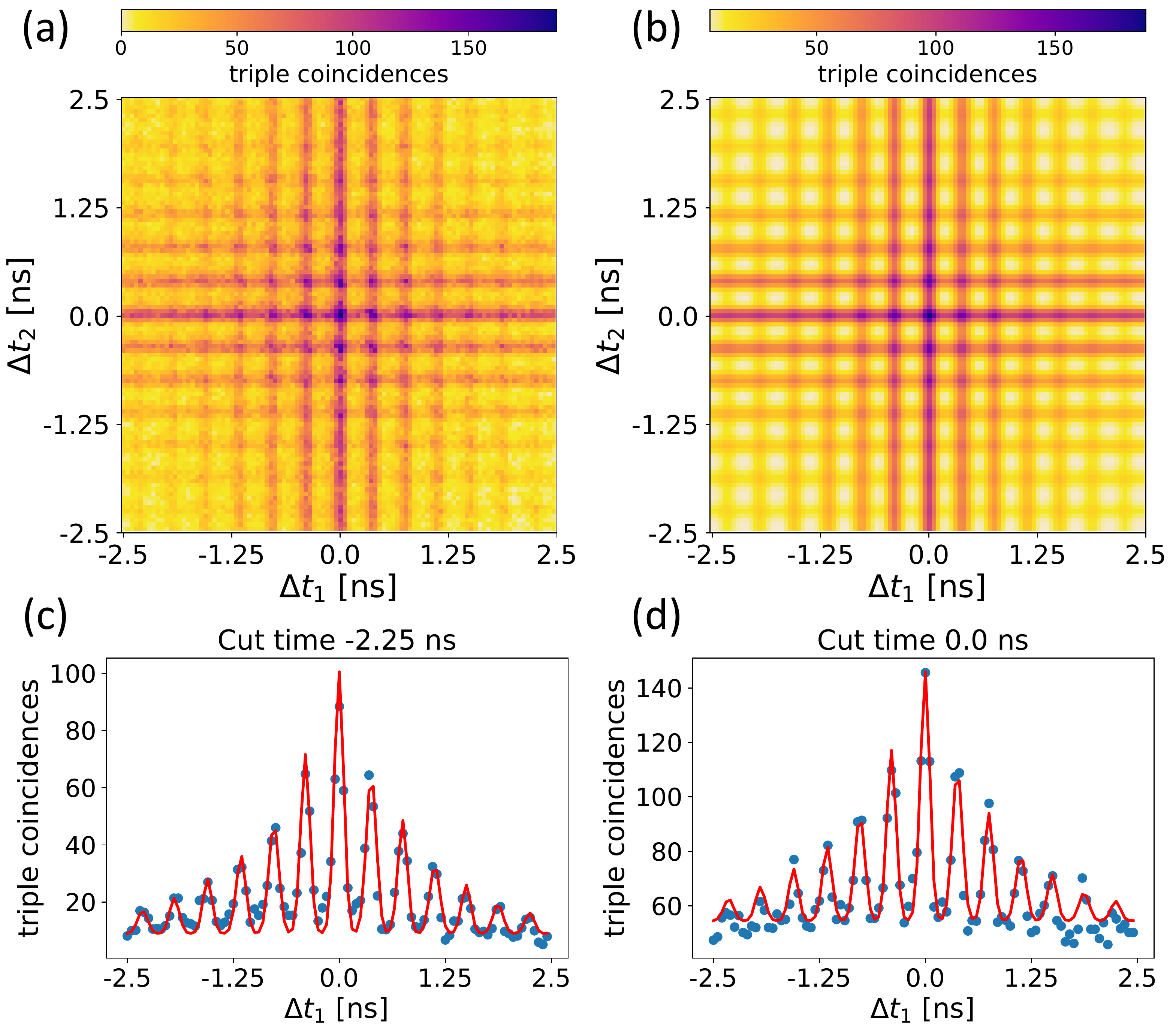}
	\caption{Comparison of measured and theoretically predicted second-order signal-signal-idler correlation with a binning of 50\,ps. (a) measured correlation function. The measurement shows separated peaks with high triple coincidence probability and stripes from uncorrelated photons. (b) theoretical correlation function which is identical to Fig.~\ref{fig:dnum600p600_1200_summation_resolution_10ps_avby10} but binned to a resolution of 50\,ps to match the measurement and convoluted with an overall detector resolution of 125\,ps. (c) Comparison of the measured (dots) and theoretical model (solid line) at $\Delta t_1=-2.25$\,ns. To compensate for the noise in the measurement, the average value over 5 bins are used for the direct comparison. (d) the same as (c) for $\Delta t_1=0$\,ns. }
		\label{fig:50ps_binning_theory_and_measurement}
\end{figure}

Fig.~\ref{fig:50ps_binning_theory_and_measurement}a shows a magnified part of the second-order signal-signal-idler correlation measurement for comparison with a convolution of the theoretical expectation with an overall temporal resolution of the setup of 125\,ps (see Fig.~\ref{fig:50ps_binning_theory_and_measurement}b). A 2D cut through the measured data and the theoretical expectation at $\Delta t_1=-2.25$\,ns shows that theory and measurement match very well (see Fig.~\ref{fig:50ps_binning_theory_and_measurement}c). Another interesting cut is through the middle ($\Delta t_1=0$) with the highest coincidence probability as shown in Fig.~\ref{fig:50ps_binning_theory_and_measurement}d. It can be seen that the theory describes the measurement very precisely.

\section{Conclusion}
We presented a complete and generalized theoretical description of the signal-idler, signal-signal and signal-signal-idler correlation for a continuously pumped cavity-enhanced SPDC. The theoretical results match our measurements exactly. Therewith, we demonstrate that the time correlation of a cavity-enhanced SPDC source can directly be derived from the joint-spectral density, which is much easier to access experimentally than correlation measurements with a very high temporal resolution. Our theoretical description can be used to design cavities which fit to the requirements of an application even when special temporal correlations, as in time-bin entangled sources, are needed. Furthermore, it ensures the full knowledge of a generated quantum state which is advantageous for testing physical concepts or improving the accuracy of measurements.
\begin{acknowledgments}
This work was supported by the German Research Foundation (DFG) Collaborative Research Center (CRC) SFB~787 project C2 and the German Federal Ministry of Education and Research (BMBF) with the project Q.Link.X. 
\end{acknowledgments}

\bibliography{paper_bib}

\end{document}